\documentclass[12pt,twoside]{article}

\usepackage{asp2006}
\usepackage{epsf}
\usepackage{epsfig}
\usepackage{lscape}
\usepackage{graphicx}

\newcommand{\msun}{\ensuremath{M_{\odot}}}

\begin{document}

\title{Broad-line AGNs with Candidate Intermediate-mass Black holes in the
Sloan Digital Sky Survey}   
\author{Xiaobo Dong\altaffilmark{1}, Tinggui Wang\altaffilmark{1},
Weimin Yuan\altaffilmark{2}, Hongyan Zhou\altaffilmark{1}, Hongguang
Shan\altaffilmark{2}, Huiyuan Wang\altaffilmark{1}, Honglin
Lu\altaffilmark{1}, and Kai Zhang\altaffilmark{1} }

\affil{}    
\altaffiltext{1}{Center for Astrophysics, University of Science and
Technology of China, Hefei, Anhui, 230026, China;
xbdong,twang@ustc.edu.cn} \altaffiltext{2}{National Astronomical
Observatories/Yunnan Observatory, Chinese Academy of Sciences, P.O.
Box 110, Kunming, Yunnan 650011, China; wmy@ynao.ac.cn}

\begin{abstract} 
We have conducted a systematic search of AGNs with IMBHs from the
SDSS DR4. As results we found 245 candidates of broad-line AGN with
M$_{BH}<10^6$\msun~estimated from the luminosity and width of the
broad H$\alpha$ component. Compared to the pioneer Greene \& Ho (2004)
sample of 19 IMBH AGNs, our sample
has improved in covering a larger
range of the Eddington ratio, as well as
black hole mass  and redshift,
taking the advantage of our AGN-galaxy spectral
decomposition algorithm.
Among these, thirty-six have $L_{bol}/L_{Edd} < 0.1$,
hinting that a significant fraction of IMBHs might exist with weak
or no nuclear activity.
\end{abstract}


\section{Introduction}
Astrophysical black holes with masses in the $10^{3-6}$\msun~range
(intermediate-mass black hole, IMBH) are of significant implications
to black hole research and cosmology. The search for IMBHs turns out
to be a difficult task, however, since IMBHs are beyond the reach of
direct dynamical measurement. The most promising approach is to
search for dwarf AGNs hosted in small galaxies. So far, convincing
evidence for the existence of IMBHs has been found in only a dozen
AGNs (NGC\,4395, Filippenko \& Ho 2003; POX\,52, Barth et al. 2004;
Greene \& Ho 2006; SDSS\,J1605+1748, Dong et al. 2007). Greene \& Ho
(2004, hereafter GH04) have conducted a pioneering systematic search
for IMBH AGNs from the SDSS DR1 over a sky area of 1360 deg$^2$,
yielding 19 candidates that are preferentially accreting close to
the Eddington limit (mainly due to their selection method). A
fundamental question remains open whether IMBHs are truly rare or
they are just mostly quiescent.

To address this and other related questions we have conducted a
thorough search for IMBH AGNs using data from the SDSS DR4.

\section{Results}   
We performed spectral analysis for all the galaxies and QSOs with
redshift $z \leq 0.35$ from the SDSS DR4 spectroscopic database
covering 4783 deg$^2$. Starlight and nuclear continuum were
subtracted from SDSS spectra using the method as described in Zhou
et al.\ (2006)\footnote{We employ the Ensemble Learning Independent
Component Analysis algorithm to model starlight (see Lu et al. 2006
for details).}. Emission line spectra were fitted using the code as
described in Dong et al.\ (2005). Based on fitted line parameters,
we culled broad-line AGNs following the criterion as described in
Dong et al.\ (2005). Black hole masses were estimated from the
luminosity and width of a broad H$\alpha$ component (Greene \& Ho
2005; Greene \& Ho 2006). Finally, we found 245 broad-line AGN with candidate IMBHs,
sampling the ranges of $5 \times 10^4 < M_{BH} < 10^6$\msun, $0.006
< z < 0.3$ and $0.02 < L_{bol}/L_{Edd} < 8$ (See Figure 1).

As noted by GH04, for IMBH AGNs whose spectrum are dominated by
narrow emission lines or which are embedded in a much more luminous
host galaxy, weak broad lines are hard to discern; this is
especially true for those in a low accretion state. For the recovery
of weak broad lines, accurate decomposition of AGN--galaxy spectra
is essential, as well as careful fitting of the lines. Resulting
from our algorithms improved in both aspects, we found 36 candidate
IMBHs with $L_{bol}/L_{Edd} < 0.1$. Figure 2 shows such an example,
SDSS\,J112628.73+472453.8 ($z=0.033$) that has $M_{BH} \approx 8
\times 10^5$\msun~and $L_{bol}/L_{Edd} \approx 0.03$. Considering
the selection effects that is seriously against finding such
objects, we suggest that there might exist a large fraction of IMBHs
with weak or no AGN activity. This can be examined in future via
deriving the $L_{bol}/L_{Edd}$ distribution of IMBH AGNs (Dong et
al., in prep.) using Monte-Carlo simulations to handle the various
selection effects (the SDSS magnitude limit, starlight dilution, and
the broad-line criterion, etc.).


\acknowledgements 
This work is supported by Chinese NSF grants NSF-10533050 and
NSF-10573015, and the BaiRenJiHua program of the CAS.  X.\,Dong is
partially supported by a postdoc grant from Wang Kuan-Cheng
Foundation.


\begin{figure}[tbp]
\plotone{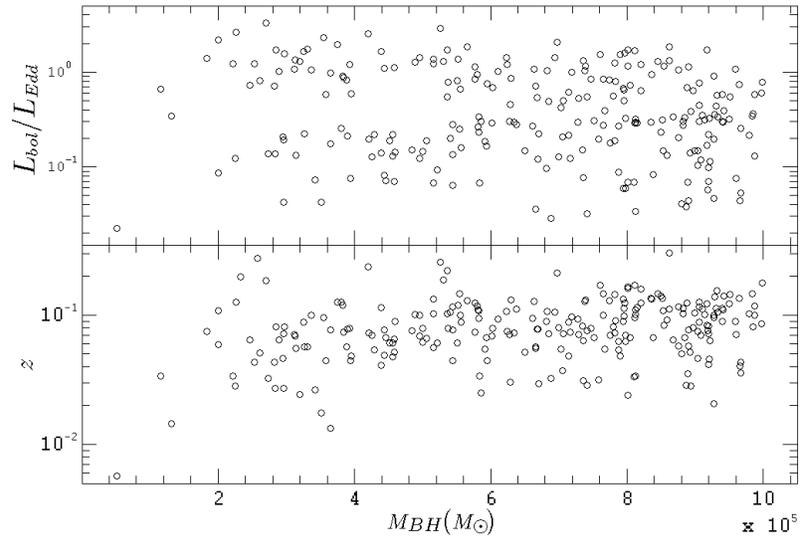} \caption{Distribution of black hole mass,
redshift, and Eddington ratio for the 245 objects of our IMBH AGN
sample drawn from SDSS DR4.}
\end{figure}

\begin{figure}[tbp]
\plotone{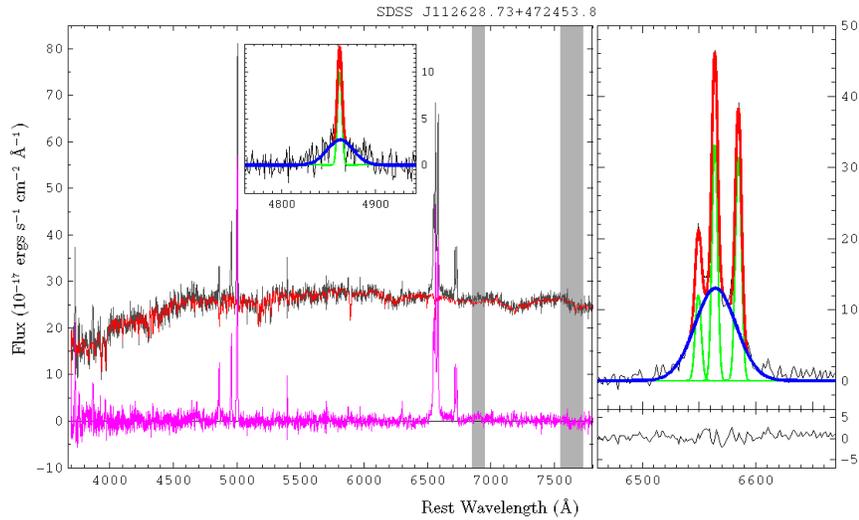}\caption{Demonstration of the discovering
optical spectra for an IMBH AGN with $L_{bol}/L_{Edd} \approx 0.03$.
Left: The SDSS spectrum (gray), the model (red), and the
starlight/continuum subtracted residual (magenta). The gray regions
are contaminated by telluric features and hence masked out during
the fit. The inset shows the narrow/broad decomposition of H$\beta$
profile. Right: The line decomposition in the H$\alpha +$[N II]
region. The upper panel shows the original data (gray line), the
fitted narrow lines (green lines), the fitted broad H$\alpha$ (blue
thick line) and the sum of the fit (red line); the lower panel, the
residual of the fit.}
\end{figure}

\end{document}